\documentclass[prl,twocolumn,aps,showpacs,superscriptaddress]{revtex4}
\usepackage{graphics}
\usepackage{epsfig}
\usepackage{amsmath}
\usepackage{amssymb}
\usepackage{bm}
\usepackage{color}

\begin{document}
\title{Flow-induced Agitations Create a Granular Fluid}


\author{Kiri Nichol}
\affiliation{Kamerlingh Onnes Lab, Universiteit Leiden, Postbus
9504, 2300 RA Leiden, The Netherlands}

\author{Alexey Zanin}
\affiliation{Kamerlingh Onnes Lab, Universiteit Leiden, Postbus
9504, 2300 RA Leiden, The Netherlands}

\author{Renaud Bastien}
\affiliation{Kamerlingh Onnes Lab, Universiteit Leiden, Postbus
9504, 2300 RA Leiden, The Netherlands}

\author{Elie Wandersman}
\affiliation{Kamerlingh Onnes Lab, Universiteit Leiden, Postbus
9504, 2300 RA Leiden, The Netherlands}

\author{Martin van Hecke}
\affiliation{Kamerlingh Onnes Lab, Universiteit Leiden, Postbus
9504, 2300 RA Leiden, The Netherlands}

\date{\today}

\begin{abstract}
We fluidize a granular medium through localized stirring and probe
the mechanical response of quiescent regions far away from the
main flow. In these regions the material behaves like a liquid:
high-density probes sink, low-density probes float at the depth
given by Archimedes' law, and drag forces on moving probes scale
linearly with the velocity. The fluid-like character of the
material is set by agitations generated in the stirred region,
suggesting a non-local rheology: the relation between applied
stress and observed strain rate in one location depends on the
strain rate in another location.

\end{abstract}

\maketitle

What governs the flow of granular media? At the grain level,
interactions are mediated by  collisions and contacts
\cite{review}. While rapid flows where collisions dominate can
be described by advanced kinetic theories \cite{goldhirsch},
and the understanding of flows where both contacts and
collisions are important has recently advanced tremendously
\cite{gdr,pouliquennature}, it remains difficult to describe
slow flows where enduring contacts dominate the interactions.

Aspects of such slow grain flows can be captured by a frictional
rheology in which the friction laws acting at the grain scale are
translated to effective friction laws for the stresses acting at a
coarse-grained level \cite{nederman,behringernature,unger}. In
such a Mohr-Coulomb picture, granular media remain jammed when the
ratio of shear $\tau$ to normal stresses $P$ is below a critical
value given by an effective friction coefficient, $\mu$, while
slowly flowing grains correspond to stresses close to the yielding
criterion: $\tau/P \approx \mu$.

This framework is, however, not complete. The combination of
rate independence and a sharp yielding criterium leads to a
description which predicts the localization of flows in shear
bands of vanishing width and a corresponding sharp separation
between stationary zones and flowing zones \cite{nederman}.
However, in experiments shear bands are found to be of finite
width and the boundary between flowing and stationary zones is
not sharp, with creep flow occurring even far away from the
main shear band
\cite{komatsu,losertpre,2008review,fenistein,cheng,mueth}. The
first key question is therefore: what is the nature of the
nearly-stationary zones far away from the main flow? A second
key question is motivated by the observation that, for slow
flows, the flow rate is independent of the stresses. But if the
flow rate is not determined by the stresses what then is the
physical mechanism that sets the flow rate of slow granular
flows? \cite{nederman,gdr,behringernature}.

\begin{figure}[tbp]
\centering
\includegraphics[clip,width=1\linewidth]{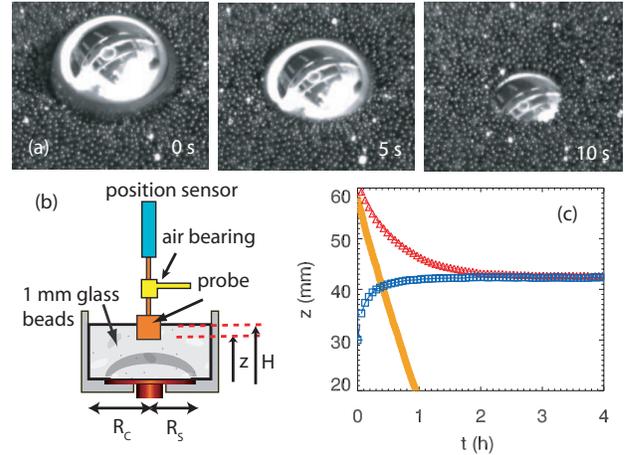} 
\caption{(Color online) (a) Snapshots of a stainless steel ball of
diameter 25 mm and mass 64 g, sinking into a stationary granular
fluid. The granular fluid is generated in a split-bottom shear
cell (inner radius $R_c=81$ mm, disk radius $R_s=60$ mm) filled to
a height of $H=60$ mm with millimetric glass beads and driven at a
rate $\Omega=0.1$ rps. (b) Experimental setup, with the dome-like
shear zone indicated in dark grey. (c) Examples of probe position,
$z$, as function of time, $t$, for a filling height of $H=60$ mm.
A high-density object sinks in the sand (orange $M=48$ g, $D=18$
mm, $\Omega=3\times 10^{-3}$ rps). A low-density object ($M=40$ g,
$D=40$ mm, $\Omega=0.1$ rps) sinks until reaching an equilibrium
depth (red)
--- and rises up to this same depth if initially deeply submerged
(blue).} \label{fig:setup}
\end{figure}

Here we address these questions by locally stirring a container of
glass beads at a rate $\Omega$ while probing the mechanical
response of the essentially quiescent regions near the surface,
away from the shear band (Fig.~1a-b). We find that a heavy object
--- such as a steel ball --- sinks slowly into the sand with a
rate proportional to $\Omega$ (Fig.~1a). Moreover, low density
objects sink (or rise) to the depth predicted by Archimedes' law
(Fig.~1c). Therefore, granular materials do not exhibit a yield
stress in the presence of flow: {\em flow fluidizes granular
media}. By observing the motion of probes immersed in the sand, we
find that the drag forces acting on the probes are linear in
velocity. This suggest that the material is viscous, although we
find that the viscosity depends strongly on filling height,
location and mass of the probes --- the rheology of the material
is highly nonlinear. Moreover, the relation between applied stress
and observed strain rate in one location depends on the strain
rate in another location \cite{boqcuetnature,bocquetprl}. Our
findings highlight novel behavior which we believe to be crucial
for the development of better models of slow granular flows
\cite{depken,pouliquenmodel,bocquetprl}.

{\em Setup ---} Our experiment consists of a split-bottom shear
cell \cite{fenistein,cheng,depken,unger}, filled with glass
beads of mean diameter $d=1$ mm to a height $H$ 
(Fig.~\ref{fig:setup}). The bottom disk (radius $R_S= 60$ mm)
is driven by a micro-stepper motor at a rate, $\Omega$, which
ranges from $10^{-4}$ to $1$ rps. The relative humidity of the
system is controlled at
$8\pm2$ \% at room temperature. Prior to beginning a
measurement, the grains are stirred with a rod, the grain
surface is leveled, and the bottom disk is spun at a rate of
$0.5$ rps for 20 s. However, we have found no evidence for
systematic dependence of the long-time behavior of the probe on
the preparation history. Grooves and dimples are machined into
the surface of the container and the disk to create a rough
boundary and the resulting grain flows have been studied
extensively \cite{fenistein,cheng,depken,unger}.

In order to investigate the liquid-like properties of the
system, smooth aluminum cylinders of mass $M$ and diameter $D$
are immersed in the grains. These probes are attached to a
shaft that passes through an air-bearing, which fixes the
horizontal position of the probe while allowing the probe to
rotate and to move in the vertical direction. The probe shaft
next passes into a LVDT position sensor (DC Fastar 2M, accuracy
of 2 $\mu m$) which we employ to measure the position of the
bottom surface of the probe, $z$.

\begin{figure}[tbp]
\centering
\includegraphics[clip,width=1\linewidth]{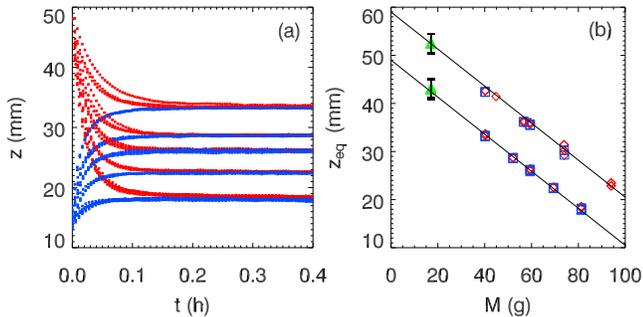}
\caption{(Color online) (a) Evolution of the probe position ($z$)
for $\Omega=0.1$ rps, $H=50$ mm, $D=40$ mm and five different
masses ranging from 40 to 81 g. (b) The equilibrium depths,
$z_{eq}$, as a function of probe mass for $H=50, 60$ mm  and
$D=40$ mm. The red diamonds and blue squares correspond to the
sinking and rising cases as shown in panel (a). The green
triangles correspond to lighter probes detached from the inductor
measurement device and imaged with a CCD camera. The straight
lines show Archimedes' law for a density of 1.92 kg/L and finite
size correction factor $\kappa=1.08$ --- the respective effective
heights, $H-\delta H$, are 49 and 59 mm.} \label{fig:eqdepth}
\end{figure}

{\em Phenomenology ---} For the large filling heights ($H/R_s >
0.8$) examined here, the shear zone that emanates from the edges
of the disk is submerged in the bulk of the material (Fig.~1b),
and the residual flow observed at the free surface is several
orders of magnitude smaller than the driving rate $\Omega$
\cite{fenistein,cheng}. A probe placed at the surface of the beads
will get stuck at a small depth if the disk is not rotating (for
example, for $D=30$ mm and $M=100$ gr, the initial depth is less
than 5 mm). However, when the disk begins rotating the probe will
immediately start to sink into the grains. Heavy probes will
continue to sink until they reach the bottom of the container,
while light objects reach a well-defined floating depth (Fig.~
\ref{fig:setup}c). By observing the motion of such probes we will
address the following questions: What sets the floating depth of
light objects? Does the material exhibit a yield stress? What
governs the drag forces on the probes?

{\em Archimedes' law ---} As shown in Fig.~\ref{fig:eqdepth}a, we
have determined the floating depths ($z_{eq}$) for a series of
probes of varying mass. For each mass, three sinking and three
rising experiments are conducted --- in all cases the equilibrium
depths, $z_{eq}$, are consistent within 0.6 mm. We conclude that
there is no appreciable yield stress in the material: if the
system did exhibit a yield stress, the sinking and rising probes
would reach different equilibrium positions. As shown in
Fig.~\ref{fig:eqdepth}b, the equilibrium depth varies linearly
with $M$. Both this linear dependence and the convergence of the
rising and sinking probes to the same floating depth are
suggestive of Archimedes' law, which in this case reads:
\begin{equation}\label{arch}
z_{eq} = H(z_{eq})-\delta H-\frac{4M}{\pi (\tilde{D})^2 \rho} =
H-\delta H -\frac{4M}{ \pi D^2 \kappa\rho}  ~.
\end{equation}
Here, both $\tilde{D}$ and $\kappa$ are effective parameters
that incorporate finite size effects. First, the probe
displaces grains as it sinks, so the actual height of the
grains, $H(z)$, depends on the probe position. Second, the
finite size of the grains suggests that the effective size of
the probe, $\tilde{D}$, may be somewhat larger than the real
value: $D<\tilde{D}<D+d$. Both effects can be incorporated in
the finite size parameter $\kappa$
--- the first effect is of order $1+(D/(2R_c))^2 \approx 1.06$,
the second effect is of order $(\tilde{D}/D)^2$ which ranges
from 1 to 1.05 --- hence we estimate $\kappa$ to be between
1.06 and 1.11. Finally, we allow for a small correction,
$\delta H$, which is on the order of a grain size and takes
into account the lower packing density of grains near the
bottom of the probe. Direct measurement of the mass and volume
of well-compacted grains yields a density of $1.92 \pm 0.05$
kg/L.

As shown in Fig.~\ref{fig:eqdepth}b, the equilibrium depth is well
described by equation (\ref{arch}) for $\kappa=1.08$ and $\delta H
= d$, where $d$ is the diameter of the beads. The finite size
correction, $\kappa$, lies within the expected range, and the
height correction $\delta H$ is also small
--- we thus conclude that Archimedes' law describes the floating
depths of our probes accurately, provided that finite size
corrections are properly taken into account. Note that
Archimedes' law has also been observed in a granular system in
which the lateral boundaries are vibrated
 \cite{xxx} --- but this driving appears far more vigorous
than in our system.

\begin{figure}[tbp]
\centering
\includegraphics[clip,width=1\linewidth]{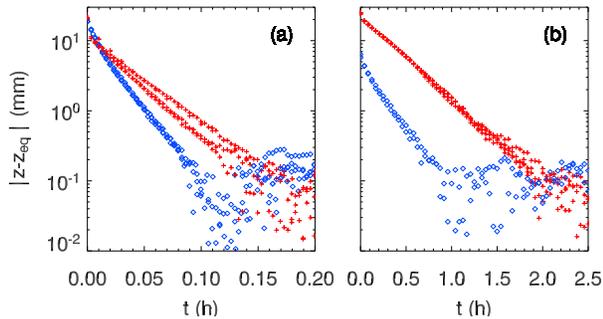}\\
\caption{(Color online) Exponential approach to equilibrium depth
as demonstrated by sinking probes (red crosses) and rising probes (blue circles)
for $M=40$ g, $D=40$ mm, $\Omega=0.1$ rps and $H=50$ mm (a) and
$H=60$ mm (b).\label{fig:eqplot}}
\end{figure}

{\em Viscous Force ---} As shown in Fig.~\ref{fig:eqplot}, the
probe position approaches the equilibrium depth exponentially
for small $|z-z_{eq}|$ \cite{notenoise}. The data suggests that
the characteristic times for rising probes are somewhat smaller
than that for sinking probes, although the apparent slope of
the curves is very sensitive to the precise choice of $z_{eq}$.
In the remainder of this paper we focus primarily on sinking
probes where we can vary the probe mass over a wide range.

Since the difference between the gravitational and buoyant
forces is linear in $z-z_{eq}$, the exponential relaxation
implies that the drag force ($F_d$) on a probe is proportional
to its velocity. Ignoring geometrical form factors for drag on
a cylinder, we define the effective viscosity, $\eta$, of the
granular fluid using $F_d \!=\! - \eta D dz/dt$. Combining this
drag force with the buoyant force, $F_b$, and the gravitational
force suggested by Archimedes' law yields the following
equation of motion for the probes:
\begin{equation}\label{eq:forces}
\eta {D} \frac{dz}{dt}   =  -M g + F_b ~.
\end{equation}
In contrast, drag forces on intruders in quiescent or oscillated
granular media reported before \cite{geng,candelier} exhibit a
finite threshold.

In the remainder of this paper we will address the following two
questions. {\em (i)} How does the effective viscosity depend on
the parameters $H$ and $M$? {\em{(ii)}} Is the viscosity set by
the local residual flow near the probe?

\begin{figure}[tbp]
\centering
\includegraphics[clip,width=1.0\linewidth]{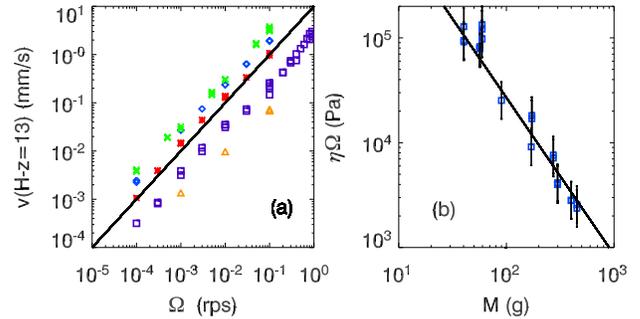}\\
\caption{(Color online) (a) Probe velocity 13 mm below the surface
as a function of disk rotation speed $\Omega$. Red stars:
$H\!=\!60$ mm, $M\!=\! 77$ g, $D \!=\!18 $ mm, Blue diamonds:
$H\!=\!50$ mm, $M\!=\!46$ g, $D\!=\!18$ mm, Purple squares:
$H\!=\!60$ mm, $M\!=\!48$ g, $D\!=\!18$ mm, Green crosses:
$H\!=\!60$ mm, $M\!=\!86$ g, $D\!=\!18$ mm. Orange triangles:
$H\!=\!60$ mm, $M\!=\!46$ g, $D\!=\!18$ mm with a probe immersed
at a radius of $15$ mm away from the center. The line has slope 1.
(b) Rescaled viscosity, $\Omega \eta$, at $z=50$ mm as a function
of probe mass for $D\!=\!40$ mm, $H\!=\!60$ mm. The line is a
guide to the eye corresponding to a power law with exponent -1.5.
}\label{fig:mass_omega_scaling}
\end{figure}

{\em Viscosity ---} From a wide range of experimental data we
conclude that the viscosity is inversely proportional to
$\Omega$. In Fig.~\ref{fig:mass_omega_scaling}a we plot the
immersion speed, $v$, at a fixed depth ($z=H-13\mbox{ mm}$) as
a function of $\Omega$ for a number of probes. Clearly, $v
\propto \Omega$ over a wide range of disk rotation speeds.
The characteristic time scale of the exponential relaxation of
floating probes is also proportional to $\Omega$ (not shown).
Hence, the relevant timescale for probe motion is set by
$\Omega$, and internal timescales (such as given by vibrations)
appear irrelevant.

In a true liquid, the viscosity is independent of the mass of an
object sinking in the system. However, the viscosities of our
granular liquids exhibit a surprising dependence on the probe mass
(Fig. \ref{fig:mass_omega_scaling}b). Moreover, we found that the
precise form of $\eta(M)$ depends strongly on $H$ and weakly on
the measurement depth, $z$. The detailed rheology is complicated
and we leave a detailed study to further work.

\begin{figure}[btp]
\centering
\includegraphics[clip,width=1.0\linewidth]{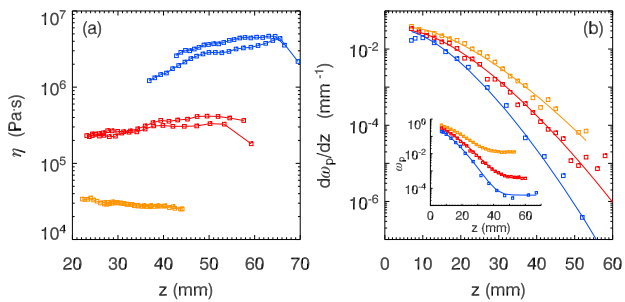}
\caption{(Color online) (a) Viscosity as a function of probe
position for a probe with $M\!=\!59$ g, $D\!=\!30$ mm and
$H\!=\!50,60$ and $70$ mm (lower orange curve, middle red curve
and upper blue curve respectively). (b) The variation of the
strain rate $\partial_z \omega_p$ with $z$ for $H\!=\!50,60$ and
$70$ mm. Inset: $\omega_p$ as a function of the height above the
spinning disk. Curves are of the form $\omega_p(z) - \omega_p(z=H)
\sim \exp(-(z/\xi)^{1.5})$, with $\xi=9.5, 11.5$ and $13$ mm
respectively.} \label{fig:viscscaling_precessionspeed}
\end{figure}

In Fig.~\ref{fig:viscscaling_precessionspeed}a we plot the
measured viscosity for a single probe ($M\!=\!59$ g, $D\!=\!30$
mm) as a function of $z$ for three values of $H$ (50, 60 and 70
mm). These viscosities are obtained by measuring $z(t)$ and
then {invoking} Eq.~(\ref{eq:forces}) to calculate the
effective viscosity. From
Fig.~\ref{fig:viscscaling_precessionspeed}a it is apparent that
the viscosity depends on the filling height, $H$. Moreover, the
viscosity for a given filling height changes very little with
the probe depth, which is consistent with our observation that
the probe position of light probes evolves exponentially over a
substantial range. Clearly, the strong dependence on $H$ and
the weak dependence on $z$ rules out a picture in which the
viscosity simply depends on the distance to the flowing zone.

{\em Nonlocal Flow Rule ---} Even though the flow and strain
rates near the free surface are very small, they are not zero
\cite{fenistein,cheng,unger}. It is therefore instructive to
ask if the probe motion is determined by the local residual
flow near the probe. To answer this question, we will compare
the $H$ and $z$ dependence of the viscosity, shown in
Fig.~\ref{fig:viscscaling_precessionspeed}a, to the dominant
local strain rate. For dome-like shear bands, the strain in the
region above the center of the disk is torsional \cite{cheng},
and so the strain rate varies as $(\partial_z \omega|_{r=0})$.

We have measured the precession rate
$\omega_p(z):=\omega(z)|_{r=0}/\Omega$ by inserting tiny
vane-like probes in the center of the grain flow and observing
the rotation of the probe with a rheometer. The resulting
precession rates, $\omega_p(z)$, are shown in
Fig.~\ref{fig:viscscaling_precessionspeed}b for $H=50,60$ and
$70$ mm, and are in good qualitative agreement with earlier MRI
measurements and simulations \cite{cheng}. Our larger
measurement range allows us to establish that $\omega_p(z)$
approaches the surface precession rate, $\omega_p(z=H)$, faster
than an exponential but slower than a Gaussian, and our results
are fitted well by an expression of the form $\omega_p(z) -
\omega_p(z=H) = \omega_p(z=0) \exp(-(z/\xi)^{1.5})$, where
$\omega_p(z=0)$ captures slip near the bottom disc, and the
characteristic length scale $\xi$ is of order 10 mm.

By either differentiating this expression, or by numerically
differentiating the measured data, we can determine $\partial_z
\omega_p(z)$: the result is shown in
Fig.~\ref{fig:viscscaling_precessionspeed}b. Clearly the local
strain rate and $\eta(z)$ correlate very poorly
--- while the former changes over four decades for $H=70$ mm,
the latter changes over less than half a decade.

{\em Conclusion and Outlook ---} By applying localized shear to
a container of glass beads we have created a granular fluid
with unusual properties. The first surprise is that our
granular fluid does not exhibit the glassy behavior typical of
agitated granular media
\cite{tap,umbanhowar,dauchotglass,durianair}, and that the
characteristic time scale is simply set by the inverse driving
rate $1/\Omega$. We speculate that the application of steady
shear does not allow for caging and other typical glass-like
effects, but instead leads to fluidization of the material.

The second surprise is that the drag forces on the intruder are
simply proportional to the velocity, which is in contrast to
earlier work \cite{candelier,geng}. The third surprise is that
our granular fluid does not exhibit a finite yield stress. The
disappearance of the yield stress in the presence of flow
suggests that the stress and strain tensor are co-linear,
consistent with recent simulations \cite{depken}.

The final surprise is that the local rheology of the material is
set by the flow in the stirred region, and {\em not} by the local,
residual (creep) flow. We suggest that random grain motion in the
flowing zone leads to grain agitations even far away from the
flow, and that these agitations generate the liquid-like behavior.
Hence, the rheology is nonlocal: the relation between applied
stress and observed strain rate in one location depends on the
strain rate in another location. Whether this is similar to
observations of nonlocal rheology recently observed in emulsions
\cite{boqcuetnature} is at present an open question.

{\em Acknowledgments ---} We thank J. Mesman for technical
assistance and M. Cates for discussions.


\begin{thebibliography}{21}
\bibitem{notenoise} The noisy behavior seen close to the equilibrium
depth is not due to measurement noise but rather to the random
motion that the floating probe experiences (of order of 1/10th of
a bead diameter or less).
\end{thebibliography}
\end{document}